\documentclass[11pt]{article}

\usepackage[preprint]{acl}

\usepackage{times}
\usepackage{latexsym}
\usepackage{enumitem}
\usepackage{algorithm}
\usepackage{algpseudocode}
\usepackage[most]{tcolorbox}
\usepackage{xcolor}
\usepackage{caption}

\usepackage{amsmath}
\usepackage{amssymb} 
\usepackage{booktabs}             
\usepackage{multirow}      \usepackage{makecell}       
\usepackage{graphicx}            
\usepackage{colortbl}       \usepackage{array}      
\usepackage{xcolor}    \usepackage{float}   \usepackage{subcaption}
\usepackage{enumitem}

\definecolor{highlight}{HTML}{D9EAF7}

\usepackage{makecell}
\usepackage{pifont} % for ding symbols
\newcommand{\cmark}{\ding{51}} % ✓
\newcommand{\xmark}{\ding{55}} % ✗
\definecolor{rowhl}{RGB}{221, 235, 247} 

\usepackage[T1]{fontenc}
\usepackage[utf8]{inputenc}

\usepackage{microtype}

\usepackage{inconsolata}

\usepackage{graphicx}

\newcommand{\modelname}{\textnormal{MACA}}

\title{Multi-Agent Coordination Adaptation via Structure-Guided Orchestration}

\author{
Haoran Li\textsuperscript{1,*},
Shulun Chen\textsuperscript{2,*,\dag},
Shaoyuan Sun\textsuperscript{3},
Hanchen Wang\textsuperscript{2}
\\
\textsuperscript{1}Nanjing University
\\
\textsuperscript{2}University of Technology Sydney, Sydney, Australia
\\
\textsuperscript{3}University of New South Wales, Sydney, Australia
\\
}

\begin{document}
\maketitle
\begin{abstract}
As large language model (LLM)-based multi-agent systems scale to handle increasingly complex tasks, balancing structural stability and dynamic adaptability becomes increasingly challenging. Existing systems typically adopt either structure-centric methods, committing to structures determined upfront that limit fine-grained control, or orchestration-centric methods, adapting decisions dynamically while leaving coordination structure implicit and unstable. To address this challenge, we revisit multi-agent coordination from a probabilistic perspective, casting it as posterior inference over the joint distribution of structure and orchestration. We introduce $\modelname$, an automated coordination framework that learns a task- and budget-conditioned structural prior over agent participation and interactions. This prior guides a policy-based orchestration as an approximation to posterior inference, enabling efficient solutions with fine-grained control. Across benchmarks, $\modelname$ outperforms adaptive multi-agent baselines by an average of 8.42\% while using 43.19\% fewer tokens. Further investigation reveals that joint adaptation of structure and orchestration suppresses redundant interactions, converging coordination toward task-effective execution. The code is available at: \href{https://github.com/However-Li/Multi-Agent-Coordination-Adaptation-via-Structure-Guided-Orchestration}{\texttt{https://github.com/However-Li/MACA}}. 
\end{abstract}

\renewcommand{\thefootnote}{\fnsymbol{footnote}}

\footnotetext[1]{Equal Contribution.}
\footnotetext[2]{Corresponding Author: Shulun.Chen@student.uts.edu.au}

\vspace{-4mm}
\section{Introduction}
\vspace{-2mm}
\label{Intro}
In an era marked by the maturation of foundation models~\cite{tu2024overview}, computational resources~\cite{burns2016borg}, and low-latency communication~\cite{adhikari20226g}, machine cognition is undergoing a paradigm shift from isolated computation to collective intelligence. As Minsky envisioned in Society of Mind~\cite{minsky1986society}, intelligence arises from the interplay of simple agents, and this vision now unfolds at scale. 

Within this context, LLM-based multi-agent systems~\cite{he2025llm,li2024survey,cheng2024exploring} focus on harnessing collective intelligence through coordinated reasoning and organization to address increasingly complex challenges that transcend individual capability. Such LLM-based multi-agent systems have been demonstrated to be effective across a broad range of application domains, including question answering~\cite{chen2025multi,zhu2024autotqa,zhang2024chain}, software development~\cite{zhang2024aflow,wang2025megaagent}, and data analysis~\cite{xiao2024cellagent,rasheed2024can,wang2025large}, where structure and orchestration become critical for effective problem solving~\cite{wu2023autogen}.

\begin{figure}[!t]
    \centering   \includegraphics[width=0.48\textwidth]
    {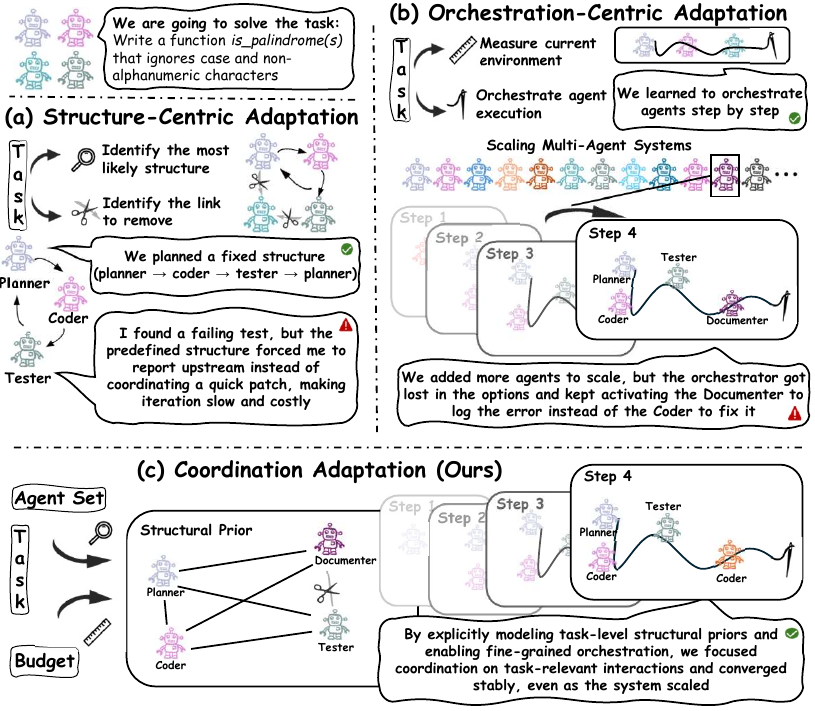}
    \vspace{-13pt}    
    \caption{(a)~\textbf{Structure-centric methods} fix a topology before inference, limiting adaptation as task states evolve. (b)~\textbf{Orchestration-centric methods} offer step-by-step adaptability without a predefined structure, but incur high-variance coordination as scale increases. (c)~\textbf{Our method} bridges these two paradigms by guiding adaptive orchestration with an explicit structural prior, achieving both stability and flexibility.}
    \label{fig:1}
    \vspace{-20pt} 
\end{figure}

Early approaches to LLM-based multi-agent coordination, such as ChatDev~\cite{qian2024chatdev}, MetaGPT~\cite{hong2023metagpt}, and AgentVerse~\cite{chen2024agentverse}, typically rely on hand-crafted topologies and fixed interaction orders~\cite{guo2024large}. While effective at encoding domain-specific structure~\cite{yang2025topological} and enabling role-level functional specialization~\cite{lin2025creativity,naik2025agentmisalignment}, such designs rely heavily on manually engineered agent or coordination rules~\cite{tang2024medagents}, and typically assume a fixed collaboration structure shared across tasks, prompting efforts for autonomous multi-agent systems. \citet{zhang2024cut} and ~\citet{talebirad2023multi} have formalized multi-agent systems as computational graphs, enabling structural and communication optimization. Methods such as GPTSwarm~\cite{zhuge2024gptswarm} and G-Designer~\cite{zhang2024g} investigate learning adaptive communication structures or interaction patterns to reduce redundancy. A closer examination of recent autonomous mechanisms reveals two predominant paradigms, as illustrated in Figure~\ref{fig:1}: \textbf{(I) Structure-Centric Adaptation:} Given a query, the system adapts the multi-agent topology either by explicitly inferring a task-specific interaction structure~\cite{zhang2025multi,yuan2025evoagent,shang2024agentsquare}, or by refining a pre-defined structure via agent generation~\cite{tian2025agentinit}, agent selection~\cite{zhang2024g}, agent dropout~\cite{wang2025agentdropout}, and communication pruning~\cite{zhang2024cut}. Since the system commits to a query-conditioned interaction structure prior to execution, it implicitly assumes that coordination efficiency is largely determined by this structural choice, leaving limited capacity for dynamic adaptation as task states evolve. \textbf{(II) Orchestration-Centric Adaptation:} Such methods adapt system behavior through sequential decision making, typically by selecting the next agent(s) at each step. Some works rely on heuristic or rule-based strategies~\cite{rasal2024navigating,rasal2024llm,qayyum2025llm}, while more recent methods formulate orchestration as an optimization problem, leveraging reinforcement learning~(RL)~\cite{sun2024llm,dang2025multi,zhang2025osc,liu2025llm} to optimize coordination policies. Despite their practical effectiveness, the absence of explicit interaction modeling induces role drift and high-variance credit assignment under increasing task complexity.

Revisiting these paradigms reveals a fundamental limitation of current adaptive multi-agent systems: \textbf{coordination is frequently treated as a single, monolithic process}.
While~\citet{yang2025topological} highlight the role of  topological structure as a primary research objective and~\citet{bhatt2025should} further investigate when orchestration is necessary, the separation between structure and orchestration prevents the system from jointly reasoning about how coordination should be organized and evolve over time, rendering coordination brittle and inefficient under scaling complexity~\cite{cemri2025multi,zhang2025agent}. Moreover, resource budgets (e.g., token limits) do not merely act as external constraints, but fundamentally shape the feasible coordination space and its inherent performance trade-offs. Ignoring these signals during either structural formulation or dynamic orchestration leads to unstable learning dynamics.

To address the above challenges, we propose a Multi-Agent Coordination Adaptation Framework~($\modelname$) that casts multi-agent coordination as a posterior inference problem, where structural priors define a constrained interaction space and orchestration policies optimize within it. Specifically, our framework advances multi-agent coordination through two key innovations: \textbf{(I) Structural Prior Learning.} $\modelname$ models the structural prior as a task- and budget-conditioned variable by jointly estimating agent relevance and interaction plausibility. This process infers a probabilistic interaction graph that constrains agent participation and information flow, thereby yielding a principled foundation for downstream orchestration.
\textbf{(II) Token-Aware Orchestration.} Given the inferred structure, $\modelname$ performs orchestration within the constrained space. The orchestration policy operates over a reduced space, where structural priors act as constraints through reward modulation, enabling flexible adaptation as task states evolve while maintaining scalability and efficiency.

\begin{figure*}[t]
    \centering   \includegraphics[width=0.93\textwidth]{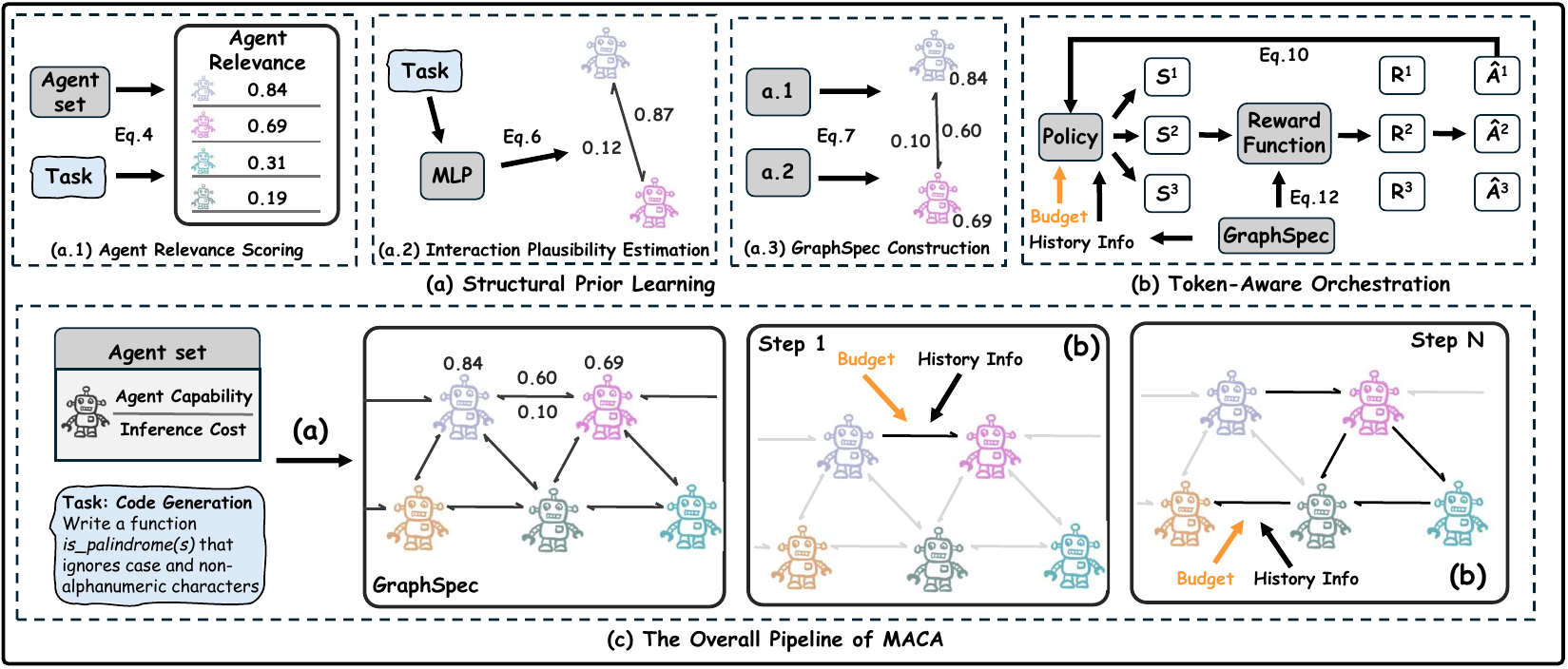}
    \caption{The overall framework of $\modelname$. It consists of three main components: (a) Structural Prior Learning, (b) Token-Aware Orchestration, and (c) the overall pipeline for a given task.}
    \label{fig:2}
    \vspace{-2mm}
\end{figure*}

Our contributions are summarized as follows: 
\begin{itemize}[leftmargin=*, nosep]
  \item We are the first to cast multi-agent optimization as posterior inference over structural and orchestration variables. This elevates adaptive coordination from architecture selection or sampling to a principled probabilistic formulation.
  \item We introduce a novel and adaptive inference framework $\modelname$, which learns a structural prior, subsequently optimizing orchestration within the generated prior to enable fine-grained and resource-efficient coordination.
  \item Extensive experiments across six benchmarks show that $\modelname$ yields robust and efficient coordination, surpassing adaptive multi-agent baselines with an average 8.42\% performance boost and a 43.19\% token reduction, thus achieving stronger task generalization and budget stability.
\end{itemize}

\section{Methodology}
\label{s3}
Figure~\ref{fig:2} provides an overview of our proposed Multi-Agent Coordination Adaptation framework, namely $\modelname$. 
We construct a structural prior from a space of agent compositions and interaction patterns. 
Building upon this prior, $\modelname$ approximates posterior inference over coordination decisions via reinforcement learning. The model leverages orchestration feedback to progressively refine system behaviors.
In the following sections, Section~\ref{Preliminaries} formally defines the coordination formulation and optimization objective of our framework,
Section~\ref{Prior} presents the construction of structural prior,
and Section~\ref{Orchestration} details the token-aware orchestration for adaptive multi-agent systems.

\subsection{Preliminaries}
\label{Preliminaries}
\noindent\textbf{(a) System Definition.}
We introduce the first unified definition for LLM-based multi-agent coordination systems, formulated as 
$
\mathcal{M} = \{\mathcal{G}, \mathcal{T}\}
$, where coordination is characterized by structure $\mathcal{G}$ and orchestration $\mathcal{T}$. 
Specifically, the structure is defined as a directed graph
$
\mathcal{G} = (\mathcal{V}, \mathcal{E}),
$
where \(\mathcal{V} \subseteq \mathcal{O}\) is a subset of a universal agent set \(\mathcal{O}\), each node $v_i \in \mathcal{V}$ represents an agent as in previous practices~\cite{zhang2024cut}. \(\mathcal{E} \subseteq \mathcal{V} \times \mathcal{V}\) specifies admissible interaction relations among agents, each edge $e_{ij} \in \mathcal{E}$ characterizes the potential for
interaction from agent $v_i$ to agent $v_j$,
thereby defining a structured interaction space for downstream
orchestration.
The process
$\mathcal{T}$
represents an ordered sequence of orchestration decisions.  A particular realization $\tau$ can be modeled as a state-action trajectory
$\tau = (s_0, a_0, s_1, a_1, \ldots, s_T)$,
where each state $s_t$ summarizes the coordination context at step $t$,
including the task input, intermediate agent outputs, and available
resource signals.
Each action $a_t$ corresponds to an orchestration decision,
such as selecting a subset of agents to invoke, routing information
among agents, or terminating execution.
The action selection process is guided by the structure $G$,
i.e., $a_t \sim \pi(\cdot \mid s_t; G)$.

\noindent\textbf{(b) Posterior Coordination.}
We refine the coordination mechanism to a posterior inference problem.
Rather than treating the structure $\mathcal{G}$ and the orchestration $\mathcal{T}$ as fixed design choices,
we model them as random variables that govern the solution process.
Given a task instance $x$ and a resource budget $b$, we model a hierarchical dependency.
A structure $G$ is first drawn from a task- and budget-conditioned structural space.
This structure constrains the distribution of the execution orchestration $\tau$.
The resulting orchestration then determines the fidelity of producing the optimal output $y^\star$.
Accordingly, coordination reduces to finding the joint posterior over structure and trajectory that best
explains $y^\star$:

\begin{small}
\vspace{-3mm}
\begin{equation}
\begin{aligned}
\label{eq:1}
p(G,\tau \mid x,b,y^\star)
\\
\;\propto\; &
\underbrace{p(y^\star \mid \tau, x)}_{\textsc{Output Fidelity}}
\cdot
\underbrace{p(\tau \mid G, x, b)}_{\textsc{Orchestration}}
\cdot
\underbrace{p(G \mid x, b)}_{\textsc{Structural Prior}}
\end{aligned}
\end{equation}
\end{small}

\noindent The posterior factorizes into (I) a structural prior $p(G\mid x,b)$ that assigns probability to
valid agent topologies under the given conditions, (II) an orchestration $p(\tau\mid G,x,b)$ capturing how execution evolves within the structural constraints, and (III) output fidelity
$p(y^\star\mid \tau,x)$ that measures how likely an orchestration yields the optimal solution.

\noindent\textbf{(c) Problem Formulation.} To compute the intractable posterior in Eq.~\ref{eq:1} and thereby infer the joint distribution of structure $G$ and orchestration $\tau$, we adopt variational inference~\cite{levine2018reinforcement} to approximate the target distribution $p(G, \tau \mid x, b, y^\star)$ with a parameterized generative policy $\pi_\theta(G, \tau \mid x, b)$. Specifically, the Kullback-Leibler (KL) divergence between the variational distribution and the target posterior is defined as follows:

\begin{small}
\vspace{-3mm}
\begin{equation}
\begin{aligned}
D_{\mathrm{KL}}(\pi_\theta \parallel p) \\
= \mathbb{E}_{(G, \tau) \sim \pi_\theta} &\left[ \log \pi_\theta(G, \tau \mid x, b) - \log p(G, \tau \mid x, b, y^\star) \right]
\label{eq:kl_derivation}
\end{aligned}
\end{equation}
\end{small}

\noindent By minimizing the KL divergence, the inference process can be formulated into an optimization problem. 
We define a utility function $U(\tau, y) = \log p(y \mid \tau, x)$, representing the log-likelihood of the ground truth outcome with given orchestration and task instance. 
Assuming a fixed budget constraint implicit in the structural prior, the minimization of the KL divergence is mathematically equivalent to maximizing the evidence lower bound (ELBO), which yields the optimization objective:

\begin{small}
\vspace{-3mm}
\begin{equation}
\max_{\theta}\;
\mathbb{E}_{\tau \sim \pi_\theta, G }
\Big[
U(\tau, y^\star)
+
\alpha \cdot \Omega(\pi_\theta \mid G, x, b)
\Big]
\label{eq:joint_objective}
\end{equation}
\end{small}

\noindent where the regularization term $\Omega(\pi_\theta \mid G, x, b)$ encodes an objective that constrains the orchestration policy to remain aligned with the structural prior under the task input $x$ and budget $b$.
The coefficient $\alpha \in [0,1]$ controls the trade-off between maximizing task utility and enforcing structural consistency.

\subsection{Structural Prior Learning}
\label{Prior}
Inspired by the success of learned priors for capturing data-driven regularities~\cite{ulyanov2018deep,wang2025aigc}, we introduce a structural prior termed \textbf{GraphSpec}, which models coordination uncertainty by jointly inferring \textbf{Agent Relevance} and \textbf{Interaction Plausibility}. 

% Before observing any task-specific execution policy, this prior captures intrinsic coordination regularities of the system. GraphSpec steers the orchestration space toward coherent configurations, ensuring efficient and scalable execution. 
% Specifically, we first formulate agent relevance and then model interaction plausibility to jointly construct GraphSpec.        

\noindent\textbf{Agent Relevance Scoring.} Agent suitability varies with task semantics and resource budgets. 
Recent findings~\cite{yang2025bamas,wang2024rethinking} show that complex reasoning benefits from selective agent engagement. 
Accordingly, agent relevance is estimated by modeling the semantic compatibility between each agent representation and the task–budget context.
For each agent $v_i\in\mathcal{O}$, we introduce a continuous variable $z_i\in[0,1]$ that represents its participation strength in the coordination structure. Let $
s_i =\mathrm{cos}~\!\big(\mathbf{e}(x,b),\,\mathbf{e}_i\big)
$ denote the resulting relevance score, where $\mathbf{e}(x,b)$ denotes the embedding of the input context and $\mathbf{e}_i$ represents the embedding of agent $v_i$, both obtained via a sentence encoder~\cite{bge_embedding}. $\mathrm{cos}~(\cdot,\cdot)$ denotes the cosine similarity. We then obtain a participation weight:

\begin{small}
\vspace{-1mm}
\begin{equation}
\label{eq:5}
q_i = \sigma\!\left(\frac{s_i}{\beta(b)}\right), \qquad
z_i =
\begin{cases}
q_i, & q_i \ge \gamma, \\
0,   & q_i < \gamma,
\end{cases}
\end{equation}
\end{small}

\noindent where $\beta(b)$ is a budget-dependent temperature. $\{q_i\}_{v_i\in\mathcal{O}}$ represents a continuous score. We apply a thresholded gating operation with parameter $\gamma$ to suppress low-confidence agents.
Agents with $q_i < \gamma$ are filtered out, while agents exceeding the threshold retain their participation strength. The resulting $Z_{prior}=\{z_i\}$ specifies agent relevance.

% $\mathbf{Z} = \{z_i\}$ is subsequently used to adjust inter-agent probabilities.

\noindent\textbf{Interaction Plausibility Estimation.} We model interaction plausibility as an edge-level prior over directed agent transitions $v_i\rightarrow v_j$.
Intuitively, interactions that remain consistently effective under stochastic structural variations are more likely to reflect stable coordination patterns and thus should receive higher prior probability. 
We define a policy $\pi_\phi$ that directly parameterizes graph topology via learnable edge logits $\phi = \{\ell_{ij} \in \mathbb{R}\}$, which quantify the interaction plausibility between agents. This policy independently samples directed edges $e_{ij} \sim \text{Bernoulli}(\sigma(\ell_{ij}))$. The parameters $\phi$ are optimized by minimizing the following loss function:

% The prior is iteratively refined over discrete steps $k$, with the logit at step $k$ denoted as $\ell_{ij}^{(k)}$. Starting from an uninformative initialization $\ell_{ij}^{(0)} = 0$,

\begin{small}
\vspace{-2mm}
\begin{equation}
\label{eq:6}
\mathcal{L}(\phi) = -  \mathcal{U} \cdot \sum_{(v_i,v_j) \in \mathcal{O}} \log \sigma(\ell_{ij}) + \lambda \mathcal{R}_{s}
\end{equation}
\end{small}

\noindent where $\mathcal{U} \in \{0,1\}\ $represents a utility signal. $\mathcal{R}_{s}=\frac{1}{|\mathcal{O}|}\sum_{(v_i,v_j)\in\mathcal{O}}\sigma(\ell_{ij})$ serves as a regularizer to prevent over-fitting. 
Subsequently, we maintain a buffer of the high quality interaction trajectories sampled from $\pi_\phi$, filtering for instances that achieve correct reasoning. 
These filtered trajectories are used as pseudo-labels to train a Multi-Layer Perceptron (MLP): 

\begin{small}
\begin{equation}
\label{eq:7}
    P(v_i \rightarrow v_j | x) =  \text{MLP}_\psi([\mathbf{e}_i, \mathbf{e}_j, \mathbf{e}(x)])
\end{equation}
\end{small}

\noindent The MLP maps the task $x$ to the edge interaction probability for any agent pair $(v_i, v_j)$.

% \noindent Evolving from a neutral initialization, the model distills cost-sensitive utility signals into a task-consistent prior, discovering intrinsic coordination regularities without explicit structural annotations.

\noindent\textbf{GraphSpec Construction.} We synthesize the estimated agent relevance and interaction plausibility into GraphSpec, a unified probabilistic prior that adapts the structure to the task-budget context. We modulate the raw edge probabilities by the participation strengths of the incident nodes. The effective connection probability $P_{prior}$ is defined as:

\begin{small}
\vspace{-4mm}
\begin{equation}
\tilde{p}_{ij} = p_{ij} \cdot q_j,\quad P_{prior} = [\tilde{p}_{ij}] \in [0,1]^{N \times N}
\label{eq:graphspec}
\end{equation}
\vspace{-4mm}
\end{small}

\noindent This modulation suppresses links involving low-confidence agents. The resulting GraphSpec, denoted by
$
\mathcal{GS}(x,b) = (Z_{prior}, P_{prior}),
$
defines a parameterization of the structural prior
\(p(\mathcal{G}\mid x,b)\).

\subsection{Token-Aware Orchestration}
\label{Orchestration}
% Orchestration is formulated as a sequential decision process, where a policy $\pi_{\theta}$ dynamically navigates the constrained interaction space. 
% Orchestration is performed through a policy optimization approach.

The orchestration problem is formulated as a Markov Decision Process~(MDP), and the policy is optimized using a GRPO-based method~\cite{guo2025deepseek}. 
At each step $t$, the state $s_t \in \mathcal{S}$ encapsulates the task input $x$, history state $h_t$, and budget $b_t$. The objective is to learn a policy
$\pi_\theta(a_t \mid s_t)$
that maximizes the expected return:

\begin{small}
\vspace{-2mm}
\begin{equation}
\max_{\theta}\;
\mathbb{E}_{\tau \sim \pi_\theta}
\Big[
R
-
\alpha \cdot
D_{\mathrm{KL}}
\big(
\pi_\theta
\;\|\;
\pi_{\text{ref}}
\big)
\Big]
\label{eq:grpo_objective}
\end{equation}
\end{small}

\noindent where $R=\sum_{t=1}^{T} r_t'$ denotes the total return of trajectory $\tau$, with \(r_t'\) defined in Eq.~\eqref{eq:12}. $\pi_{\mathrm{ref}}$ denotes the frozen reference policy, and $\alpha$ is a trade-off parameter. The policy $\pi_\theta$ selects an action $a_t$ from an action space $\mathcal{A} = \mathcal{O} \cup \{\text{STOP}\}$.
For each task, we sample a group of $K$ trajectories
$\{\tau^{(k)}\}_{k=1}^{K}$ using the current policy.
Let $R^{(k)}$ denote the reward of trajectory $k$.
Group-relative advantages are computed as:

\begin{small}
\vspace{-5mm}
\begin{equation}
\label{eq:group_adv}
A^{(k)}
=
R^{(k)}
-
\frac{1}{K}\sum_{k'=1}^{K} R^{(k')},
\quad
\tilde{A}^{(k)}
=
\frac{A^{(k)}}{\mathrm{Std}(\{R^{(k')}\}) + \epsilon}
\end{equation}
\end{small}

\noindent The overall loss function combines a clipped term $L_{\text{clip}}(\theta)$ with a KL divergence penalty, denoted as:

% \begin{small}
% \vspace{-4mm}
% \begin{equation}
% L_{\text{clip}}
% (\theta) = \min \left( \rho_t(\theta) \tilde{A}, \text{clip}(\rho_t(\theta), 1-\epsilon, 1+\epsilon) \tilde{A} \right), 

% \tag{10}
% \makeatletter
% \def\@currentlabel{10}
% \makeatother
% \label{eq10}
% \mathcal{J}(\theta) = -\mathbb{E} \Bigg[ \frac{1}{K} \sum_{k} \bigg( L_{\text{clip}}^{(k)}(\theta) - \alpha \cdot D_{\text{KL}}(\pi_\theta \| \pi_{\text{ref}}) \bigg) \Bigg]
% \end{equation}
% \end{small}

\begin{small}
\vspace{-4mm}
\begin{align}
L_{\text{clip}}(\theta)
&= \min \left(
\rho_t(\theta)\tilde{A},
\text{clip}(\rho_t(\theta), 1-\epsilon, 1+\epsilon)
\tilde{A}
\right),
\nonumber
\\
\mathcal{J}(\theta)
&= -
\mathbb{E} \Bigg[
\frac{1}{K}
\sum_{k}
\bigg(
L_{\text{clip}}^{(k)}(\theta)
-
\alpha \cdot
D_{\text{KL}}
\left(
\pi_{\theta}
\, \| \,
\pi_{\text{ref}}
\right)
\bigg)
\Bigg].
\tag{10}
\label{eq10}
\end{align}
\end{small}

\noindent  where $\rho_t(\theta)=\exp(\log \pi_\theta - \log \pi_{\theta_{\mathrm{old}}})$
is the likelihood ratio,
$\epsilon$ is the clipping range. A KL divergence term is introduced to regularize $\pi_{\theta}$, keeping it close to $\pi_{\mathrm{ref}}$, which balances reward maximization with stability. However, relying solely on standard regularization leaves the vast multi-agent interaction space unconstrained, allowing the policy to waste tokens on spurious interactions. To address this, we introduce a hybrid supervision mechanism that incorporates both hard constraints and soft regularization into the orchestration. We first impose constraints on the action space to filter out structurally implausible interactions. 
Given the structural mask $\mathcal{H}$, we enforce the validity of the policy $\pi_\theta$ by masking out invalid actions. The masked policy distribution is formally defined as:

\begin{small}
\vspace{-4mm}
\begin{equation}
    \pi_{masked}(a_t | s_t) = \frac{\exp(l(a_t|s_t)) \cdot \mathcal{H}_{ a_t}}{\sum_{a' \in \mathcal{A}} \exp(l(a'|s_t)) \cdot \mathcal{H}_{a'}}
    \tag{11}
    \label{eq:action_masking}
\end{equation}
\end{small}

\noindent where $l(\cdot|s_t)$ denotes the raw logits generated by the policy network. By setting the probability
of masked actions to zero, we restrict the policy to a prior-consistent subset of actions, constraining exploration to plausible connections.

To prevent the policy from engaging in aimless exploration among valid but low-value interactions, we incorporate the edge probabilities from the prior as a reference distribution $\pi_{\mathrm{mix}} =\tfrac{1}{2}\Bigl( \pi_{\mathrm{ref}}+\pi_{\mathrm{prior}}\Bigr)$. Here, $\pi_{\mathrm{prior}}$ denotes the prior action distribution induced by GraphSpec. Distinct from GRPO, we leverage a KL penalty that encourages the policy to anchor its exploration around the learned structural prior. The reward function is formulated as:

\begin{small}
\vspace{-2mm}
\begin{equation}
    r_t' = r_t - \lambda \cdot D_{KL} \left( \pi_\theta \parallel \pi_{mix} \right),
    \tag{12}
    \label{eq:12}
\end{equation}
\end{small}

\noindent where $r_t = R_{\text{acc}} - \beta C_t^{\text{token}}$ denotes the extrinsic reward, which balances task utility $R_{\text{acc}}$ against token cost $C_t^{\text{token}}$ through the trade-off parameter $\beta$. $\lambda$ is a regularization coefficient balancing task performance and structural adherence. This soft regularization keeps the policy close to the learned prior, stabilizing the training process in complex scenarios. More details of the training algorithm can be found in Appendix~\ref{app:al}.

\section{Experiments}
\vspace{-1mm}
\begin{table*}[t]
\centering
\small
\caption{Performance comparison of baseline methods on Llama-3.1-8B. The best results are shown in \textbf{bold}, and the second-best are \underline{underlined}. Avg Cost is compared among multi-agent methods to reflect their efficiency.}
\label{tab:1}
\vspace{-6pt}
\setlength{\tabcolsep}{4.6pt}
\renewcommand{\arraystretch}{1.12}

\resizebox{0.98\textwidth}{!}{
\begin{tabular}{lcc  rr rr rr  rr  rr rr}
\toprule
\multirow{3}{*}
{\makecell[c]
{\textbf{Model:}\\\textbf{Llama-3.1-8B}}} &
\multirow{3}{*}{\makecell[c]{\textbf{Structure}\\\textbf{Adaptation}}} &
\multirow{3}{*}{\makecell[c]{\textbf{Orchestration}\\\textbf{Adaptation}}} &
\multicolumn{4}{c}{\textbf{Code Generation}} &
\multicolumn{4}{c}{\textbf{Question Answering}} &
\multicolumn{4}{c}{\textbf{Math Reasoning}} \\
\cmidrule(lr){4-7} \cmidrule(lr){8-11} \cmidrule(lr){12-15}

& & 
& \multicolumn{2}{c}{\textbf{HumanEval}}
& \multicolumn{2}{c}{\textbf{MBPP}}
& \multicolumn{2}{c}{\textbf{MMLU-Pro}}
& \multicolumn{2}{c}{\textbf{ARC-C}}
& \multicolumn{2}{c}{\textbf{
SVAMP}}
& \multicolumn{2}{c}{\textbf{GSM-Hard}} \\
\cmidrule(lr){4-5} \cmidrule(lr){6-7}
\cmidrule(lr){8-9} \cmidrule(lr){10-11}
\cmidrule(lr){12-13} \cmidrule(lr){14-15}

& & 
& \textbf{Acc (\%)} & \textbf{Avg Cost}
& \textbf{Acc (\%)} & \textbf{Avg Cost}
& \textbf{Acc (\%)} & \textbf{Avg Cost}
& \textbf{Acc (\%)} & \textbf{Avg Cost}
& \textbf{Acc (\%)} & \textbf{Avg Cost}
& \textbf{Acc (\%)} & \textbf{Avg Cost} \\
\midrule

Vanilla   & \xmark & \xmark   & 60.61& 256.6& 45.97 & 138.4 & 38.50& 246.7& 83.09 & 279.6 & 82.18 & 209.6 & 32.31 & 345.3 \\

\midrule

CoT   & \xmark & \xmark   & 56.44 & 297.9& 44.62 & 146.1 & 39.24 & 379.5& 83.17 & 443.7 & 82.51 & 267.8 & 36.36 & 387.0 \\
ComplexCoT & \xmark & \xmark   & 57.58 & 1154.4 & 44.18 & 247.4 & 38.54 & 1304.2 & 82.18 & 883.1 & 86.33 & 2658.9 & 32.83 & 2766.3 \\
Self-Refine & \xmark & \xmark   & 62.12 & 1609.8 & 47.53 & 498.7 & 33.75 & 1646.5 & 83.16 & 1191.4 & 83.67 & 1195.3 & 26.26 & 2525.5 \\
SC~(CoT$\times$5) & \xmark & \xmark   & 55.38 & 2389.6 & 46.36 & 1206.9 & 41.62 & 2406.9 & 84.16 & 1545.0 & 88.67 & 1080.7 & 40.40 & 1639.8 \\

\midrule

DyLAN      & \xmark & \xmark   & \underline{72.73} & 15256.7 & 12.50 & 11647.8 & 47.26 & 11874.6 & 83.33 & 3471.6 & 86.33 & 6675.6 & 33.63 & 15338.0 \\
MacNet     & \xmark & \xmark  & 68.18 & 8782.6 & 47.83 & 6794.8 & 25.47 & 10284.4 & 83.83 & 3596.3 & 79.74& 6912.6 & 19.84 & 12468.3 \\
AgentVerse & \xmark & \xmark  & 66.18 & 6826.9 & \underline{48.94} & 5017.5 & 41.69 & 6193.2 & 85.61 & 3014.9 & 86.03 & 4184.7 & 38.19 & 6768.4 \\
\midrule

AgentPrune   & \cmark & \xmark & 62.50 & \underline{2357.0} & 31.44& \underline{2787.8} & 43.53& \underline{2713.6}& \underline{86.09}& 2691.3& 88.44 & 
\underline{2780.8} & 33.86 & 3815.8 \\

MaAS         & \cmark & \xmark & 65.15 & 4796.9 & 42.05 & 3215.7 &  43.94&   2978.4& 43.83 & \underline{2674.7} & 92.28 & 2827.7 & \textbf{51.52} & \underline{2224.8} \\

Puppeteer    & \xmark & \cmark & 71.49 & 2798.3 & 47.17 & 3696.5 & \underline{51.25} & 4097.8 & 85.57 & 3753.3 & \underline{94.18} & 4204.7 & 49.25 & 4196.6  \\

\midrule
\rowcolor{rowhl}

\textbf{$\modelname$ (Ours)} & \cmark & \cmark & \textbf{75.76} & \textbf{2100.1}  & \textbf{49.23} & \textbf{2412.3} & \textbf{52.67} & \textbf{2117.8} & \textbf{87.75} & \textbf{1656.3} & \textbf{96.00} & \textbf{2057.2}& \underline{50.30} & \textbf{1602.5} \\
\bottomrule
\end{tabular}}
\vspace{-2mm}
\end{table*}

\subsection{Experiment Setup}
\noindent\textbf{Benchmarks and Metrics.}
We comprehensively evaluate $\modelname$ across six benchmarks spanning three domains.
\textbf{(I) Code generation}, HumanEval~\cite{chen2021evaluating} and MBPP~\cite{austin2021program};
\textbf{(II) Question Answering}, MMLU-Pro~\cite{wang2024mmlu}
and ARC-C ~\cite{allenai:arc};
and \textbf{(III) Math Reasoning}, SVAMP~\cite{patel2021nlp} and GSM-Hard~\cite{gao2022pal}.
We evaluate models in terms of \textbf{Accuracy} and \textbf{Average Cost}. Accuracy is computed as
$\frac{1}{N}\sum_{i=1}^{N}\mathbb{1}(\hat{y}_i \text{ is correct})$,
where correctness follows task-specific criteria (e.g., Pass@1 or Exact Match).
Average Cost is measured as
$\frac{1}{N}\sum_{i=1}^{N} c_i$, where $c_i$ denotes the token cost of task $i$, including both prompt tokens and completion tokens.
The dataset statistics are in
Appendix~\ref{app:dataset}.

\noindent\textbf{Baselines.} 
We compare $\modelname$ against baselines categorized by their coordination and adaptation mechanisms:
\textbf{(I) single-agent methods} including CoT~\cite{wei2022chain},
ComplexCoT~\cite{fu2022complexity}, Self-refine~\cite{madaan2023self}, and Self-Consistency~\cite{wang2022self}.
\textbf{(II) hand-crafted multi-agent systems} including DyLAN~\cite{liu2024dynamic},
AgentVerse~\cite{chen2024agentverse},
and MacNet~\cite{qian2024scaling}.
\textbf{(III) Adaptive multi-agent systems} including
AgentPrune~\cite{zhang2024cut},
MaAS~\cite{zhang2025multi} and
Puppeteer~\cite{dang2025multi}.
Additional details for baselines are provided in Appendix~\ref{app:baseline}.

\begin{table}[t]
\centering
\small
\caption{Results on Llama-3.1-70B. The best and runnerup results are bolded and underlined, respectively. Avg Cost is compared among multi-agent methods.}
\label{tab:2}
\vspace{-6pt}
\setlength{\tabcolsep}{6pt}
\renewcommand{\arraystretch}{1.15}

\resizebox{\linewidth}{!}{
\begin{tabular}{l cc cc cc}
\toprule
\multirow{2}{*}{\makecell[c]
{\textbf{Model:}\\\textbf{Llama-3.1-70B}}} &
\multicolumn{2}{c}{\textbf{MMLU-Pro}} &
\multicolumn{2}{c}{\textbf{HumanEval}} &
\multicolumn{2}{c}{\textbf{GSM-Hard}} \\ 
\cmidrule(lr){2-3} \cmidrule(lr){4-5} \cmidrule(lr){6-7} 
& \textbf{Acc (\%)} & \textbf{Avg Cost}
& \textbf{Acc (\%)} & \textbf{Avg Cost}
& \textbf{Acc (\%)} & \textbf{Avg Cost} \\ 
\midrule

Vanilla        & 53.50 & 250.6 & 78.78 & 239.8 & 48.87 & 206.7 \\ 
CoT            & 54.25 & 274.1 & 80.30 & 247.1& 57.14 & 292.8 \\

DyLAN          & \underline{60.09}& 12068.9& \underline{84.25}& 12984.5& 55.05& 14186.4\\
AgentPrune     & 57.74& 3128.7& 80.25& \underline{3186.9}& 51.44 & 4372.5 \\
MaAS           & 58.74& \underline{2868.2}& 83.33 & 3758.5 & \underline{64.03} & \underline{3125.8} \\
\midrule
\rowcolor{rowhl}

\textbf{$\modelname$ (Ours)} & \textbf{64.57}& \textbf{2438.6}& \textbf{88.89} & \textbf{2672.2} & \textbf{67.37} & \textbf{2747.7} \\
\bottomrule
\end{tabular}}
\vspace{-4mm}
\end{table}

% \begin{table}[t]
% \centering
% \small
% \caption{Results with baseline on Qwen3-14B.}
% \label{tab:2}
% \vspace{-6pt}
% \setlength{\tabcolsep}{6pt}
% \renewcommand{\arraystretch}{1.15}

% \resizebox{\linewidth}{!}{
% \begin{tabular}{l cc cc cc}
% \toprule
% \multirow{2}{*}{\makecell[c]
% {\textbf{Model:}\\\textbf{Qwen3-14B}}} &
% \multicolumn{2}{c}{\textbf{MMLU-Pro}} &
% \multicolumn{2}{c}{\textbf{HumanEval}} &
% \multicolumn{2}{c}{\textbf{GSM-Hard}} \\ 
% \cmidrule(lr){2-3} \cmidrule(lr){4-5} \cmidrule(lr){6-7} 
% & \textbf{Acc (\%)} & \textbf{Avg Cost}
% & \textbf{Acc (\%)} & \textbf{Avg Cost}
% & \textbf{Acc (\%)} & \textbf{Avg Cost} \\ 
% \midrule

% Vanilla        & 0 & 0 & 0 & 0 & 0 & 0 \\ 
% CoT            & 0 & 0 & 0 & 0 & 0 & 0 \\

% DyLAN          & 0 & 0 & 0 & 0 & 0 & 0 \\
% AgentPrune     & 0 & 0 & 0 & 0 & 0 & 0 \\
% MaAS           & 0 & 0 & 0 & 0 & 0 & 0 \\
% \midrule
% \rowcolor{rowhl}

% \textbf{$\modelname$ (Ours)} & 0 & 0 & 0 & 0 & 0 & 0 \\
% \bottomrule
% \end{tabular}}
% \end{table}

\subsection{Performance Comparison}

\noindent\textbf{MACA consistently outperforms baselines.} 
As shown in Table~\ref{tab:1}, $\modelname$    achieves the highest average accuracy of 68.62\% across six benchmarks. Compared to single-agent methods, $\modelname$ yields an average accuracy improvement of $9.19 \%\sim12.54\%$. Against adaptive multi-agent baselines, $\modelname$ delivers an 8.42\% improvement in accuracy while simultaneously reducing token costs by 36.2\% to 51.9\%. These results confirm the effectiveness and cost-efficiency of $\modelname$. Notably, this advantage remains consistent across model scales, from Llama-3.1-8B to Llama-3.1-70B.

\noindent\textbf{RQ1: Can coordination itself induce reasoning capability?}
At the Llama-3.1-8B scale~(Table~\ref{tab:1}), $\modelname$ yields clear gains over single-agent baselines, suggesting that the improvements arise from coordinated interaction. While methods such as CoT~\cite{wei2022chain}  improve reasoning by prompting models to generate explicit step-by-step rationales, they remain bounded by single-model limitations and may reinforce incorrect trajectories. By contrast, $\modelname$ mitigates the self-reinforcing error loops inherent in monolithic generation through coordination among diverse agents, achieving an average absolute gain of 13.69\% over CoT on MMLU-Pro and GSM-Hard. More importantly, cross-scale results (Tables~\ref{tab:1} and~\ref{tab:2}) show that $\modelname$ with Llama-3.1-8B can outperform a vanilla Llama-3.1-70B model on GSM-Hard~(50.30\% vs.\ 48.87\%), while remaining competitive on MMLU-Pro and HumanEval~(reaching \textasciitilde97.3\% of the vanilla 70B model's performance).

% Instead, by shaping how reasoning paths intersect, challenge, and refine one another, MACA effectively imposes a structural prior over the epistemic space, making useful collisions between reasoning trajectories more likely and enabling systematic self-correction. 

\subsection{Framework Analysis}
\noindent\textbf{Ablation Study.} Table~\ref{tab:maca_ablation} presents an ablation study on key components of $\modelname$:~(1)~w/o \(Z_{prior}\), removing agent relevance prior;~(2)~w/o \(P_{prior}\), removing interaction plausibility prior;~(3)~w/o GraphSpec, eliminating the entire structural prior; and~(4)~w/o $\pi_{\theta}$, removing the learnable policy. 

\begin{table}[H]
\centering
\small
\vspace{-2mm}
\caption{Ablation study of $\modelname$.}
\vspace{-1mm}
\label{tab:maca_ablation}
\setlength{\tabcolsep}{6pt}
\renewcommand{\arraystretch}{1.15}
\resizebox{0.99\columnwidth}{!}{
\begin{tabular}{l | cc | cc}
\toprule
\textbf{Dataset} 
& \multicolumn{2}{c|}{\textbf{ARC-C}} 
& \multicolumn{2}{c}{\textbf{GSM-Hard}} \\
\cmidrule(lr){2-3} \cmidrule(lr){4-5}

\textbf{Metric} 
& \textbf{Acc (\%)} & \textbf{Avg Cost} 
& \textbf{Acc (\%)} & \textbf{Avg Cost} \\
\midrule

$\modelname$ 
& 87.75& 1656.3& 50.30& 1602.5\\

\midrule

$\modelname$ w/o \(Z_{prior}\) 
& 82.64& 1920.4& 44.37& 1885.2\\

$\modelname$ w/o \(P_{prior}\) 
& 81.31& 1896.7& 43.92& 1830.8\\

$\modelname$ w/o GraphSpec 
& 72.35& 2285.9& 36.80& 2050.1\\

$\modelname$ w/o $\pi_{\theta}$
& 85.53 & 2109.6& 47.96& 1785.5\\
\bottomrule
\end{tabular}}
\vspace{-2mm}
\end{table}

\noindent Removing GraphSpec degenerates the system into an orchestration-centric approach. Without a structural prior to constrain the vast interaction space, performance drops most severely~(14.45\% $\downarrow$ in accuracy). Conversely, removing the policy $\pi_{\theta}$ reduces the system to a structure-centric approach. While accuracy only drops slightly, the token cost surges drastically (19.40\% $\uparrow$ in cost). This reveals a clear functional division: GraphSpec preserves reasoning fidelity by constraining the search space, while the token-aware policy $\pi_{\theta}$ optimizes efficiency. Focusing on the structural prior itself, we observe that dropping either agent relevance $Z_{prior}$ or interaction plausibility $P_{prior}$ leads to distinct accuracy degradation and increased overhead. 
This highlights that the joint effect of participating agents and their interaction patterns is essential for a robust coordination prior.

% This highlights that capturing the joint distribution of who participates and how they interact is essential for a robust coordination prior.

\begin{figure}[t]
    \centering
    \includegraphics[width=0.99\columnwidth]{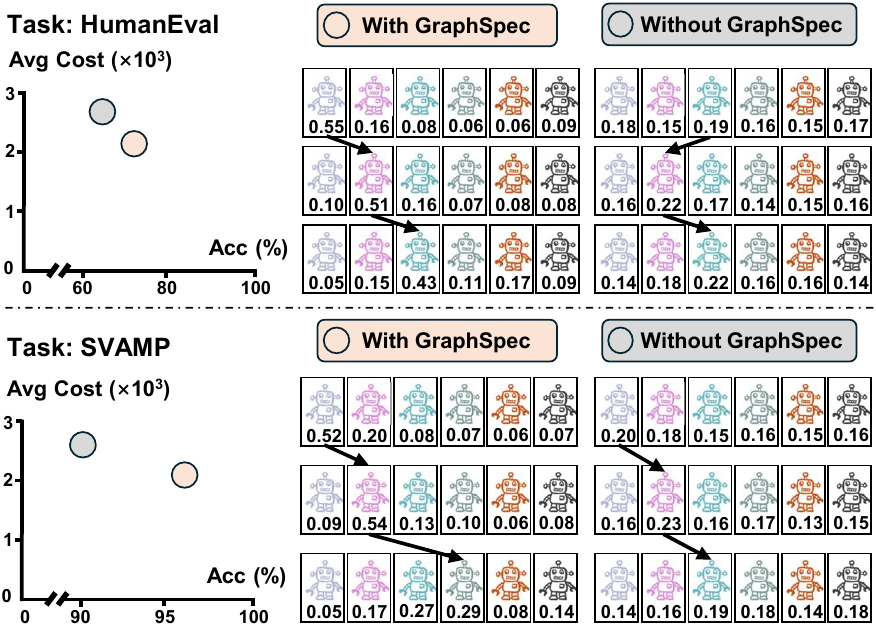}
    \caption{Impact of the prior on orchestration.}
    \label{g3}
    \vspace{-4mm}
\end{figure}

\noindent\textbf{RQ2: How does the prior affect orchestration decisions?}
Figure~\ref{g3} shows that GraphSpec imposes a structural bias on orchestration. Without it, coordination is highly unconstrained, with probability mass diffusely spread across transitions. 
With GraphSpec, the distribution concentrates on a few dominant, task-relevant paths. Even when the same agent is selected, the prior still reshapes the confidence of execution. For example, the top-2 transition mass increases from 0.37 to 0.71 on HumanEval and from 0.38 to 0.72 on SVAMP, showing that GraphSpec turns diffuse routing into concentrated, task-relevant orchestration. This concentration makes orchestration more selective and cost-efficient, improving MACA’s efficiency.

% Consequently, the prior effectively sharpens the probability distribution of agent selection, enabling the system to achieve higher accuracy with significantly reduced token consumption.

\subsection{Cost Analysis}
\begin{figure}[H]
    \centering
    \vspace{-2mm}
    \includegraphics[width=\columnwidth]{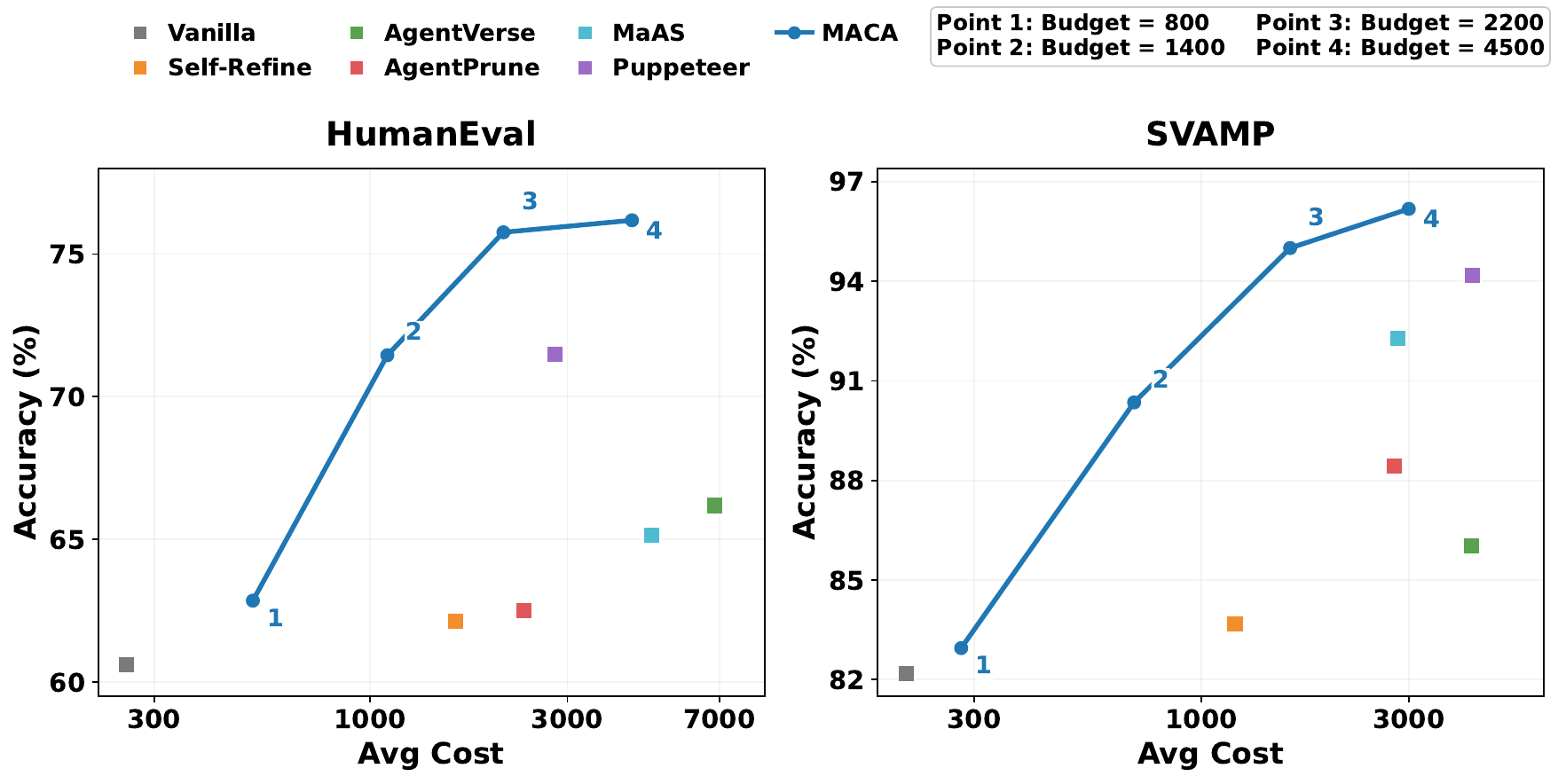}
    \caption{Cost–performance trade-offs of MACA and baseline approaches across datasets.}
    \label{g4}
    \vspace{-4mm}
\end{figure}

\textbf{RQ3: How does MACA balance cost and task performance?} 
We evaluate cost–performance by jointly considering task accuracy and token consumption during inference, as shown in Figure~\ref{g4}.    

\noindent MACA exhibits a clear cost-performance trade-off:  higher cost consistently yields higher accuracy. MACA is tunable with respect to budget, enabling practitioners to flexibly trade computation for performance under different constraints. In practice, this allows accuracy to improve in a predictable manner. More importantly, MACA achieves a more favorable cost--performance frontier than existing multi-agent baselines. This advantage stems not merely from budget tunability, but from how the budget is utilized. MACA allocates computation through structured and selective coordination, converting additional cost into accuracy gains more effectively while avoiding unnecessary interaction. 

% By penalizing excessive token consumption and encouraging efficient coordination patterns, the policy promotes concise yet effective reasoning trajectories. This leads to consistent gains in cost–performance across different task categories.

\subsection{Coordination Mechanisms Analysis}

\noindent\textbf{RQ4: What coordination patterns emerge in MACA?} Analysis of agent transition probabilities and coordination patterns on GSM-Hard trajectories reveals the following mechanisms:

\begin{figure}[H]
    \centering
    \includegraphics[width=0.93\columnwidth]{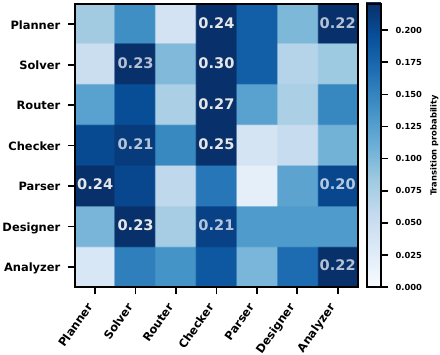}
    \caption{Column-to-row Transition Probabilities.}
    \label{g6}
    \vspace{-6mm}
\end{figure}

\noindent\textbf{Self-Correction~(X $\rightarrow$ Checker):} Figure~\ref{g6} shows a coupling between generation and verification modules, particularly transitions such as \textit{Solver} $\to$ \textit{Checker} (0.30) and \textit{Router} $\to$ \textit{Checker} (0.27), with iterative feedback (e.g., \textit{Checker} $\to$ \textit{Solver}, 0.21). This pattern mirrors the \textbf{Self-Refine}~\cite{madaan2023self}, where candidate solutions are followed by verification and revision. However, $\modelname$ goes beyond fixed self-refine loops by learning an orchestration policy over when and where verification should occur: verification is invoked more frequently along complex reasoning while being skipped for simpler steps. This mechanism allows $\modelname$ to balance accuracy and token efficiency.

\begin{figure}[t]
    \centering
    \vspace{-5mm}
    \includegraphics[width=0.89\columnwidth]{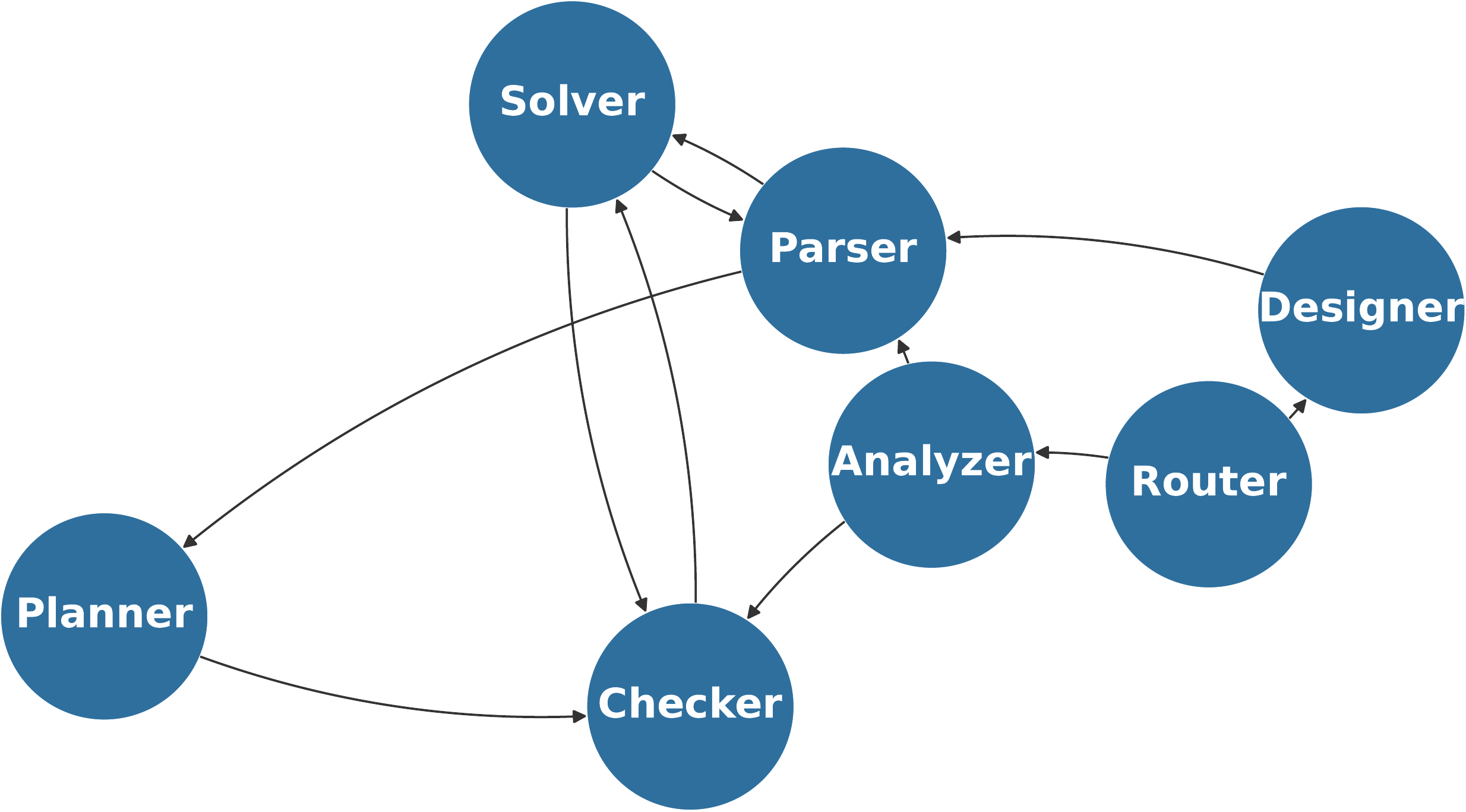}
    \caption{Dominant Three-Agent Coordination Patterns.}

    \label{g7}
    \vspace{-6mm}
\end{figure}

\noindent\textbf{Hierarchical Coordination:}
Figure~\ref{g7} aggregates the most frequent three-agent transition sequences, revealing clear hierarchical coordination patterns. The first layer, \textit{Router} and \textit{Designer}, is responsible for task decomposition and strategy selection. The second layer, \textit{Parser} and \textit{Analyzer}, translates high-level intent into structured representations. The final layer, \textit{Solver} and \textit{Checker}, carries out computational reasoning and verification. Rather than relying on manually designed workflows, these coordination strata emerge from optimizing the structural prior and orchestration policy. This process structures agent capabilities into a cascaded reasoning framework, ensuring efficient task execution.

\subsection{Sensitivity Analysis}
\begin{figure}[H]
    \centering
    \includegraphics[width=\columnwidth]{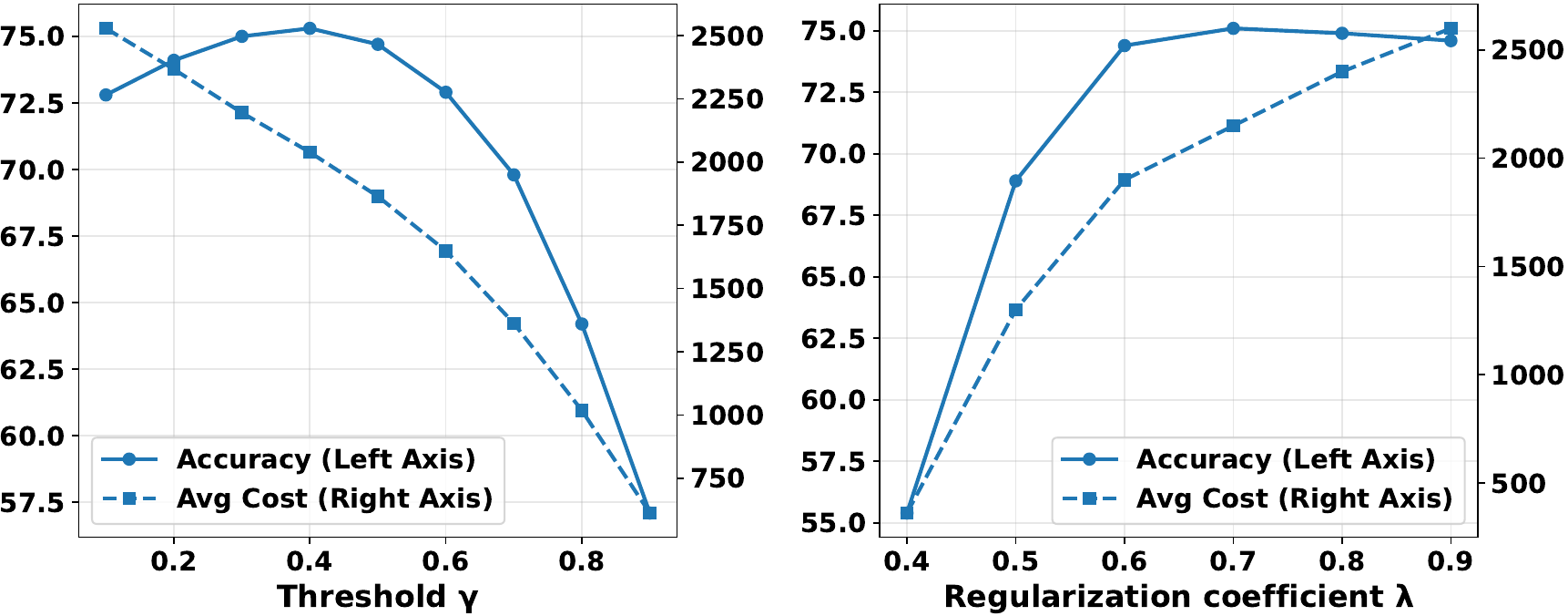}
    \caption{Parameter sensitivity analysis of MACA.}
    \label{g5}
    \vspace{-4mm}
\end{figure}

Figure~\ref{g5} illustrates the sensitivity of $\modelname$ to two core parameters: the threshold $\gamma$ in Eq.~\eqref{eq:5}, and the regularization coefficient $\lambda$ in Eq.~\eqref{eq:12}. 

\noindent\textbf{Threshold $\gamma$.} An accuracy-efficiency trade-off emerges as $\gamma$ varies. As $\gamma$ increases from 0.1 to 0.4, accuracy rises from 72.8\% to 75.2\% while token cost drops, showing that moderate thresholding removes redundant coordination. Further increasing $\gamma$ causes accuracy to decline steadily, indicating that overly aggressive sparsification discards useful coordination paths. We therefore set $\gamma=0.4$.

\noindent\textbf{Regularization coefficient $\lambda$.} Increasing $\lambda$ from 0.4 to 0.7 improves accuracy from 55.4\% to 75.1\%, but also raises token cost, indicating that stronger regularization improves orchestration quality at the expense of computation. Beyond this point, accuracy saturates while cost continues to grow, suggesting diminishing returns from overly large regularization. Accordingly, we adopt $\lambda=0.7$.

\vspace{-2mm}
\section{Related Work}

\noindent\textbf{LLM-based Multi-Agent Systems.} 
Advances in language models~\cite{guo2025deepseek,hurst2024gpt} have driven progress in LLM-based agents~\cite{zhao2024expel,li2023camel,xie2024can}. As tasks grow complex, research has increasingly turned to LLM-based multi-agent systems~\cite{bo2024reflective,du2023improving}. Prior work demonstrates empirical performance in such systems, exemplified by MacNet~\cite{qian2024scaling}, ChatDev~\cite{qian2024chatdev}, and AutoGen~\cite{wu2023autogen}. Despite progress, early approaches rely on handcrafted  structures~\cite{hong2023metagpt}, limiting the exploration of structural optimality and transferability~\cite{cemri2025multi}. Hence, research has begun to explore adaptive agentic systems.

\noindent\textbf{Adaptive Agentic Systems Optimization.} 
Recent work explores how agent compositions~\cite{zhuge2024gptswarm,zhang2024g} and execution policies~\cite{yuan2025evoagent} can be optimized, broadly categorized into two streams: (I) Structure-Centric Adaptation methods~\cite{yang2025bamas,wang2025agentdropout} adapt multi-agent topology, with methods such as AgentPrune~\cite{zhang2024cut} and MaAS~\cite{zhang2025multi} parameterizing agent interactions~\cite{li2023gslb,yan2021fp}. (II) Orchestration-Centric Adaptation methods~\cite{qayyum2025llm,rasal2024llm} introduce adaptivity at the execution level. Structure-centric methods favor stability, whereas orchestration-centric methods provide finer-grained control, yet both remain limited in isolation.

\noindent\textbf{RL as Inference.}
Reinforcement Learning~(RL) has been widely used in LLM-based multi-agent coordination~\cite{sun2024llm}. Applications include structural reasoning like BAMAS~\cite{yang2025bamas} and DyLAN~\cite{liu2024dynamic}, as well as execution orchestration such as Puppeteer~\cite{dang2025multi} and OSC~\cite{zhang2025osc}. MAPRO~\cite{zhang2025mapro} uses posterior inference for multi-agent prompt optimization. More broadly, prior research has established the concept of RL as Inference~\cite{levine2018reinforcement,o2020making,tarbouriech2023probabilistic}. This view casts control as probabilistic inference and provides a principled foundation for execution optimization.

\vspace{-2mm}
\section{Conclusion}
\vspace{-2mm}
\label{s5}
In this paper, we introduce $\modelname$, a probabilistic framework that rethinks automated multi-agent system design from a posterior inference perspective. $\modelname$ explicitly factorizes system adaptation into a task- and budget-conditioned structural prior and a token-aware orchestration policy, enabling principled uncertainty modeling and fine-grained control. $\modelname$ enables adaptive multi-agent systems that are effective across diverse tasks. The probabilistic perspective provides a foundation for future research on adaptive multi-agent systems.

\section*{Limitations}

Although $\modelname$ demonstrates clear gains in both task performance and cost-efficiency,  the current evaluation is still centered on relatively structured benchmarks, so its generalizability to more open-ended, interactive, or domain-specific settings is not yet fully established. In addition, $\modelname$ depends on a predefined agent pool with manually specified capabilities and role descriptions, which means that part of its effectiveness may still come from careful agent initialization rather than from coordination alone. Another limitation lies in the way the structural prior is learned: GraphSpec is trained from filtered high-quality trajectories and supervision signals derived from final outcome quality, which can favor coordination patterns that are easy to verify on benchmark tasks while making it harder to capture useful but less immediately rewarded interactions. This may reduce robustness in settings where credit assignment is noisy, intermediate collaboration is important, or success cannot be cleanly reflected by final answers alone.

\bibliography{reference}

\clearpage
\appendix

\section*{Appendix}

\section{LLM Usage}
ChatGPT was used solely to support language refinement during the writing process, including spellchecking, grammar improvement, and paraphrasing of the authors’ original text. The assistant was not used to generate new technical content, research ideas, or experimental results. All AI-assisted revisions were carefully checked, edited where necessary, and approved by the authors.

\section{Probabilistic Assumptions for Posterior  Inference}
We adopt a probabilistic view of coordination in which the coordination structure $G$ affects the target outcome $y^\star$ only through the induced execution trajectory $\tau$. Formally, conditioned on the input $x$ and trajectory $\tau$, the output is independent of $G$, i.e., $y^\star \perp G \mid (\tau, x)$. Under this assumption, the output fidelity term in Equation~(\ref{eq:1}) depends only on the realized trajectory and the given task, while $G$ remains important because it constrains which trajectories are plausible under the task and budget.

From this perspective, once a suitable structural prior is learned, posterior inference favors trajectories with higher utility. In $\modelname$, this intractable inference is approximated by a learnable parameterized policy, while the structural prior induces a reference distribution $\pi_{\mathrm{mix}}$ to guide exploration toward structurally plausible and high-value coordination patterns. In this sense, $\modelname$ can be interpreted as an approximation to posterior coordination inference under a learned structural prior.

\section{Experimental Details}

\subsection{Dataset}
\label{app:dataset}

Following practice in prior work~\cite{zhang2025multi,yang2025bamas}, we split each benchmark into training and test sets. As shown in Table~\ref{tab:dataset_statistics}, our experiments cover three task domains: code generation, math reasoning, and question answering. For code generation, we use HumanEval and MBPP, evaluated by pass@1. For math reasoning, we include GSM-Hard and SVAMP, both evaluated by accuracy. For question answering, we use MMLU-Pro and ARC-Challenge~(ARC-C), also measured by accuracy. 

\begin{table}[H]
\centering
\caption{Dataset Statistics.}
\label{tab:dataset_statistics}
\resizebox{\columnwidth}{!}{%
\begin{tabular}{l l c c l}
\toprule
\textbf{Domain} & \textbf{Dataset} & \textbf{\#Train} & \textbf{\#Test} & \textbf{Metric} \\
\midrule
\multirow{2}{*}{Code Generation}
& HumanEval & 96  & 68  & pass@1 \\
& MBPP      & 587 & 387 & pass@1 \\
\midrule
\multirow{2}{*}{Math Reasoning}
& GSM-Hard  & 611 & 389 & Accuracy \\
& SVAMP     & 593 & 407 & Accuracy \\
\midrule
\multirow{2}{*}{Question Answering}
& MMLU-Pro      & 731 & 487 & Accuracy \\
& ARC-Challenge & 689 & 483 & Accuracy \\
\bottomrule
\end{tabular}
}
\end{table}

\subsection{Baseline Setups}
\label{app:baseline}

To ensure fair comparison, all baselines are rerun under a unified experimental protocol. All methods use the same backbone LLM, the same dataset split, and the same evaluation metrics as in the experiments. For methods with public implementations, we follow their original core design; otherwise, we reproduce them according to the descriptions in their papers. The mean performance over three independent trials is reported. We describe the configurations of the baseline methods in detail:

\noindent \textbf{Vanilla.}
A single LLM directly produces the final answer without explicit deliberation or interaction.

\noindent \textbf{CoT.}
We use standard Chain-of-Thought prompting~\cite{wei2022chain}.

\noindent \textbf{ComplexCoT.}
We adopt complexity-based prompting~\cite{fu2022complexity} with more elaborate intermediate reasoning than standard CoT.

 \noindent\textbf{Self-Refine.}
We follow Self-Refine~\cite{madaan2023self} and let a single LLM iteratively generate, critique, and refine its answer.

\noindent\textbf{SC (CoT$\times$5).}
We apply self-consistency~\cite{wang2022self} by sampling five CoT reasoning paths and aggregating the final answer.

 \noindent\textbf{DyLAN.}
We follow the collaborative discussion setting of DyLAN~\cite{liu2024dynamic}.

 \noindent\textbf{MacNet.}
We use MacNet~\cite{qian2024scaling} with a fixed fully connected communication topology.

 \noindent\textbf{AgentVerse.}
We follow the role-based collaboration framework of AgentVerse~\cite{chen2024agentverse}.

\noindent\textbf{AgentPrune.}
We implement AgentPrune~\cite{zhang2024cut} as a structure-centric adaptive baseline that prunes agents or communication edges before execution.

 \noindent\textbf{MaAS.}
We follow MaAS~\cite{zhang2025multi} as a structure-adaptive baseline that dynamically selects task-relevant agents and interaction patterns.

\noindent\textbf{Puppeteer.}
We follow Puppeteer~\cite{dang2025multi} as an orchestration-centric baseline where a controller dynamically selects which agent to invoke at each step.

\subsection{Computational Resources}

All experiments are conducted on servers equipped with \textbf{8 NVIDIA A800 GPUs}, and mixed-precision training is used throughout. The underlying large language model is deployed through \textbf{vLLM}~\cite{kwon2023efficient} during both training and evaluation. Unless otherwise specified, all reported results are obtained under the same hardware setting.

Our framework is trained in two stages. The prior learning stage typically takes about \textbf{3--6 hours} per dataset. The policy optimization stage is more computationally intensive due to rollout sampling, reward computation, and parameter updates, and usually requires about \textbf{12--24 hours} per dataset. During evaluation, a full benchmark run typically takes about \textbf{1--3 hours}.

\newcommand{\agentbox}[3]{%
\begin{tcolorbox}[
    enhanced,
    width=\linewidth,
    colback=white,
    colframe=black!70,
    boxrule=0.8pt,
    arc=2mm,
    left=2mm,right=2mm,top=2mm,bottom=2mm,
    colbacktitle=black!75,
    coltitle=white,
    fonttitle=\bfseries,
    title=#1
]
\textbf{#1:} #2\\
\textit{Prompt summary:} #3
\end{tcolorbox}
\vspace{1mm}
}

\subsection{Agent Setups}
\label{app:agent}

We instantiate a pool of functionally specialized agents with distinct roles. 
Table~\ref{tab:agent_pools} summarizes the candidate agent pools used for different task families, while Figures~\ref{fig:shared_agents}--\ref{fig:code_agents} provide the corresponding role prompts. 

\begin{table}[H]
\centering
\scriptsize
\caption{Candidate Agent Pools.}
\label{tab:agent_pools}
\resizebox{0.98\columnwidth}{!}{
\begin{tabular}{p{0.25\columnwidth}p{0.68\columnwidth}}
\toprule
\textbf{Task Family} & \textbf{Candidate Agent Pool} \\
\midrule
\makecell[tl]{Question\\Answering~[Fig.~\ref{fig:qa_agents}]} & TaskRouter, AnalyzeAgent, ChoiceEliminator, EvidenceChecker, Skeptic, QASynthesizer \\
\midrule
\makecell[tl]{Math\\Reasoning~[Fig.~\ref{fig:math_agents}]} &
WordProblemParser, MathSolver, ArithmeticChecker, StepChecker, AlgebraSimplifier, GeneralCritic \\
\midrule
\makecell[tl]{Code\\Generation~[Fig.~\ref{fig:code_agents}]} &
AlgorithmDesigner, CodeWriting, CodeReviewer, UnitTestWriter, EdgeCaseHunter, BugFixer \\
\midrule
\makecell[tl]{Auxiliary\\Roles~[Fig.~\ref{fig:shared_agents}]} &
TaskPlanner, Summarizer, BudgetController, RedTeamCritic \\
\bottomrule
\end{tabular}}
\end{table}

\noindent In our framework, prior shaping is achieved by first estimating task- and budget-conditioned agent relevance scores to identify which roles are most useful for the current input, and then modeling interaction plausibility to characterize which inter-agent communications are likely to be beneficial. Based on these two components, the framework organizes their communication according to the learned orchestration structure. 

During analysis, we maintain a lightweight mapping that canonicalizes implementation-specific agent names into shared functional roles (e.g., mapping planning-related agents to Planner) to enable consistent role-level statistics and comparisons.

\subsection{Component and Threshold Sensitivity}
\label{app:structural-prior-threshold-sensitivity}

The structural prior in \modelname{} is not solely determined by the threshold $\gamma$.
Instead, $\gamma$ is used as a gating mechanism to filter low-confidence agents.
The full structural prior consists of two complementary components: agent relevance
$Z_{\text{prior}}$, estimated from prior knowledge, and interaction plausibility
$P_{\text{prior}}$, estimated from trajectory experience. Therefore, sensitivity
should be understood from two perspectives: component sensitivity, which studies
whether removing $Z_{\text{prior}}$ or $P_{\text{prior}}$ affects performance, and
parameter sensitivity, which studies how $\gamma$ changes the accuracy--efficiency
trade-off.

\begin{figure}[H]
\centering
\includegraphics[width=1\linewidth]{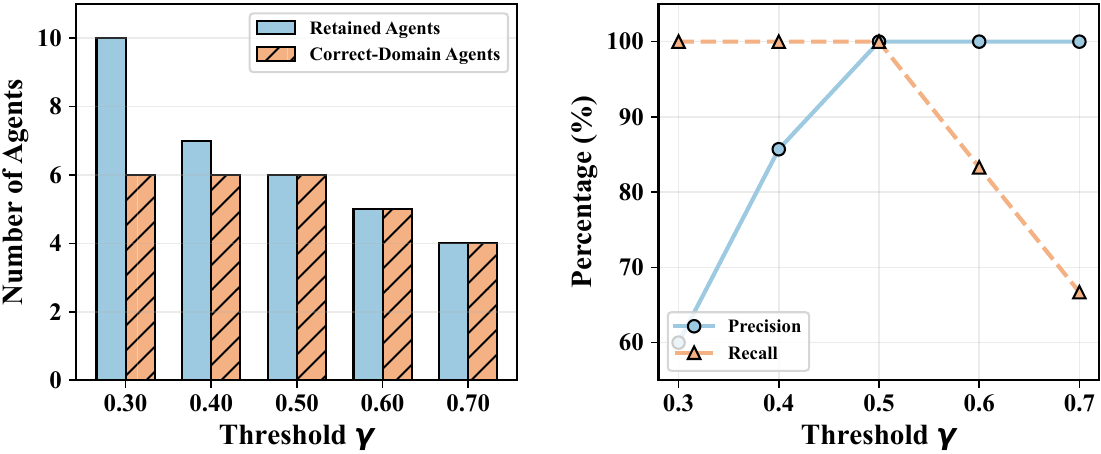}
\caption{Sensitivity of Agent-Selection Sparsity to $\gamma$.}
\label{graph_8}
\vspace{-4mm}
\end{figure}

To analyze the effect of $\gamma$ on agent-selection sparsity,  a sensitivity
study is conducted by varying $\gamma$ and measuring the precision and recall of
the retained agent set.
Here, precision denotes the proportion of retained agents that belong to the
correct domain, while recall denotes the proportion of correct-domain agents
successfully retained. As shown in Figure~\ref{graph_8},
a moderate $\gamma$ removes redundant agents while preserving the complete
correct-domain agent set. In contrast, an overly large $\gamma$ may discard useful
agents or interaction paths.

\noindent Overall, $\gamma$ controls the sparsity of the retained agent set rather than
defining the structural prior itself. Moderate values achieve a favorable
accuracy--efficiency trade-off by pruning redundant agents, whereas excessively
large values may reduce recall by filtering out useful agents.

\subsection{Algorithm}
\label{app:al}
This subsection summarizes the training procedure of $\modelname$, as detailed in Algorithm~\ref{alg:maca}. 
Training consists of two stages: learning a task- and budget-conditioned structural prior, and optimizing a token-aware orchestration policy over execution trajectories. 
We adopt this two-stage strategy to stabilize policy learning by first providing a learned structural reference distribution, rather than jointly updating the structure and policy in a non-stationary coordination space. 
The policy is trained with policy-gradient updates to favor task-effective coordination under explicit budget constraints.

\clearpage

\begin{figure*}[p]
\centering
\caption{Auxiliary role prompts.}
\label{fig:shared_agents}

\agentbox{TaskPlanner}
{Produces a coarse multi-step execution plan before detailed reasoning begins.}
{Break the task into a small number of necessary steps, identify dependencies between them, and suggest a compact execution plan.}

\agentbox{Summarizer}
{Compresses intermediate discussion into a concise state summary for downstream agents.}
{Summarize the most useful intermediate conclusions, unresolved uncertainties, and next-step recommendations in a compact form.}

\agentbox{BudgetController}
{Monitors communication cost and recommends more economical coordination behavior.}
{Estimate whether the current discussion is becoming unnecessarily expensive and suggest cheaper alternatives when possible.}

\agentbox{RedTeamCritic}
{Stress-tests the current reasoning by searching for overlooked risks, counterexamples, or failure modes.}
{Deliberately challenge the current solution, identify brittle assumptions, and point out scenarios in which the answer may fail.}

\end{figure*}

\clearpage

\begin{figure*}[p]
\centering
\caption{Question-answering role prompts.}
\label{fig:qa_agents}

\agentbox{TaskRouter}
{Determines the task type and proposes a minimal, task-appropriate coordination team.}
{Classify the task as code, math, or QA, and provide a concise routing rationale without solving the task.}

\agentbox{AnalyzeAgent}
{Extracts task constraints, key entities, hidden assumptions, and salient evidence from the input and peer outputs.}
{Analyze the problem structure carefully and summarize the most decision-relevant information.}

\agentbox{ChoiceEliminator}
{Reduces the answer space by explicitly ruling out implausible or contradicted options.}
{Eliminate incorrect choices one by one and then provide the final answer letter.}

\agentbox{EvidenceChecker}
{Verifies whether candidate answers are supported by the question facts and flags contradictions.}
{Match each option against the evidence, identify factual mismatches, and select the most supported answer.}

\agentbox{Skeptic}
{Challenges intermediate reasoning and exposes weak assumptions or unsupported jumps.}
{Critique other agents' reasoning, highlight likely errors, and question overconfident conclusions.}

\agentbox{QASynthesizer}
{Produces the final concise answer by aggregating validated evidence from upstream agents.}
{Return a short, direct final answer grounded in the strongest available evidence.}

\end{figure*}

\clearpage

\begin{figure*}[p]
\centering
\caption{Math reasoning role prompts.}
\label{fig:math_agents}

\agentbox{WordProblemParser}
{Transforms a math word problem into explicit quantities, variables, and equations before solving.}
{List known values, define variables, and derive equations or constraints without computing the final answer.}

\agentbox{MathSolver}
{Carries out the main step-by-step derivation for numerical reasoning tasks.}
{Solve the problem sequentially, show the key arithmetic or algebraic steps, and produce the final numeric answer.}

\agentbox{ArithmeticChecker}
{Recomputes arithmetic operations to detect calculation mistakes in intermediate or final steps.}
{Recalculate sums, products, ratios, and final values, then confirm or correct the numeric answer.}

\agentbox{StepChecker}
{Identifies the first invalid step in a reasoning chain and repairs it.}
{Locate the first incorrect step, explain why it is wrong, and provide a corrected derivation with unit consistency.}

\agentbox{AlgebraSimplifier}
{Simplifies symbolic expressions and rewrites equations into cleaner, easier-to-verify forms.}
{Rearrange equations and simplify intermediate expressions while preserving correctness.}

\agentbox{GeneralCritic}
{Detects hidden assumptions, missing constraints, and logically weak solution paths.}
{Review the proposed solution critically and point out logical gaps, overlooked cases, or unjustified assumptions.}

\end{figure*}

\clearpage

\begin{figure*}[p]
\centering
\caption{Code generation role prompts.}
\label{fig:code_agents}

\agentbox{AlgorithmDesigner}
{Designs the algorithmic strategy and data structures before implementation.}
{Outline the intended approach, complexity, and edge cases without writing the full program.}

\agentbox{CodeWriting}
{Produces the executable implementation for the target coding task.}
{Write the complete Python solution, preserve the required function signature, and output valid code only.}

\agentbox{CodeReviewer}
{Inspects correctness, boundary conditions, and code quality after implementation.}
{Review the candidate program for logical bugs, edge cases, and maintainability issues.}

\agentbox{UnitTestWriter}
{Constructs representative and adversarial tests for validating candidate programs.}
{Provide concise assertions or test cases that cover both standard and corner-case behavior.}

\agentbox{EdgeCaseHunter}
{Searches for pathological inputs and counterexamples likely to break a candidate solution.}
{List tricky boundary conditions or adversarial inputs and explain why they are risky.}

\agentbox{BugFixer}
{Repairs incorrect implementations using minimal but targeted modifications.}
{Identify the failure mode and provide a corrected version of the code with minimal necessary changes.}

\end{figure*}

\clearpage

\begin{algorithm*}
\caption{: Training Procedure of $\modelname$}
\label{alg:maca}
\begin{algorithmic}[1]
% Inputs cleaned up for clearer notation
\Require Training dataset $\mathcal{D}_{train}$, Agent set $\mathcal{O}$, Generation group size $g$, Reference policy $\pi_{ref}$, Budget constraint $b$, Threshold  $\gamma$, Regularization coefficient $\lambda$ 

\Statex \hspace{-1.5em}\textbf{Output}: Orchestration policy $\pi_{\theta}$

\noindent\textit{Structural Prior Learning:}

\For{each episode}
    \State Sample $\{G^{(k)}\}_{k=1}^{K} \sim \pi_\phi(\cdot \mid x)$\
    \State Update $\phi$  accordingly
     \Comment{Eq.~\ref{eq:6}}
    \State Derive $\psi$ from the sampled structures $G^{(k)}$ via $\pi_\phi$
\EndFor

\noindent\textit{Token-Aware Orchestration Policy Learning:}
\State Initialize parameters $\theta$
\For{each episode}
    \State Sample a task $x \sim \mathcal{D}_{train}$
    \State Compute agent relevance prior $Z_{prior}$ given $(x, b,\gamma)$
    \Comment{Eq.~\ref{eq:5}}

    \State Compute interaction plausibility prior $P_{prior}$ with $MLP_{\psi}$
    \Comment{Eq.~\ref{eq:graphspec}}

    \State Construct GraphSpec 
    
    \State Construct State s and Action mask $\mathcal{M}$ 
        \State Sample outputs $\{\tau^{(1)}, \tau^{(2)}, \dots, \tau^{(g)}\} \sim \pi_{\theta}(\cdot|s, \mathcal{M})$
        
        \State Obtain rewards $\{r_{\lambda}^{(1)}, r_{\lambda}^{(2)}, \dots, r_{\lambda}^{(g)}\}$\Comment{Eq.~\ref{eq:12}}
        \State Compute advantage for each group member\Comment{Eq.~\ref{eq:group_adv}}
        
        \State Compute loss w.r.t. $\pi_{\theta}$\Comment{Eq.~\ref{eq10}}
        
        \State Update $\theta$ accordingly
\EndFor
\Statex \hspace{-1.5em} \textbf{Return} $ \pi_{\theta}$
\end{algorithmic}
\end{algorithm*}

\clearpage

\section{Supplementary Results and Analysis}

\subsection{Results on Qwen2.5-14B-Instruct}
\label{app:generality-scalability}
We additionally conduct experiments using Qwen2.5-14B-Instruct as the backbone LLM. As shown in Table~\ref{tab:qwen14b_results}, \modelname{} consistently achieves the best accuracy while maintaining substantially lower token cost than dynamic multi-agent baselines. These results indicate that our proposed method is not tied to a specific LLM family, and can improve the accuracy--efficiency trade-off.

\vspace{-3mm}
\begin{table}[H]
\centering
\scriptsize
\setlength{\tabcolsep}{2.2pt}
\caption{Results on Qwen2.5-14B-Instruct.}
\label{tab:qwen14b_results}
\resizebox{\columnwidth}{!}{
\begin{tabular}{@{}lcrcrcr@{}}
\toprule
\multirow{2}{*}{\textbf{Method}}
& \multicolumn{2}{c}{\textbf{MMLU-Pro}}
& \multicolumn{2}{c}{\textbf{HumanEval}}
& \multicolumn{2}{c}{\textbf{GSM-Hard}} \\
\cmidrule(lr){2-3} \cmidrule(lr){4-5} \cmidrule(l){6-7}
& \textbf{Acc~(\%)} & \textbf{Cost}
& \textbf{Acc~(\%)} & \textbf{Cost}
& \textbf{Acc~(\%)} & \textbf{Cost} \\
\midrule
Vanilla 
& 59.12 & 248.6 
& 80.54 & 251.3 
& 47.36 & 266.8 \\

CoT 
& 60.94 & 286.7 
& 82.15 & 317.5 
& 51.28 & 391.4 \\

DyLAN 
& 64.08 & 11960.2 
& 84.37 & 12784.6 
& 52.64 & 13926.8 \\

AgentPrune 
& 63.71 & 2958.4 
& 82.76 & 3168.2 
& 55.92 & 4236.5 \\

MaAS 
& 63.22 & 3147.9 
& 85.98 & 3625.7 
& 61.37 & 3784.3 \\

\midrule
\rowcolor{rowhl}

\modelname{} 
& \textbf{64.46} & \textbf{2412.5} 
& \textbf{86.89} & \textbf{2848.9} 
& \textbf{63.28} & \textbf{2966.4} \\
\bottomrule
\end{tabular}
}
\end{table}

\subsection{Results on Reasoning-Oriented Models}
\label{results_1}

To further examine whether \modelname{} generalizes beyond instruct-tuned models, we conduct additional experiments on GSM-Hard using a reasoning-oriented 14B model.
As shown in Table~\ref{tab:reasoning_model_results}, \modelname{} consistently improves accuracy while substantially reducing token cost compared with DyLAN.
On Qwen2.5-14B-Instruct, \modelname{} improves accuracy by 10.64\% and reduces average token cost by 78.7\%.
On DeepSeek-R1-Distill-Qwen-14B, where extended reasoning introduces much higher token pressure, \modelname{} still improves accuracy by 4.03\% and reduces average token cost by 71.4\%.
These results suggest that \modelname{} remains effective under reasoning-oriented backbones.

\begin{table}[H]
\centering
\footnotesize
\setlength{\tabcolsep}{2.5pt}
\caption{Results on GSM-Hard.}
\label{tab:reasoning_model_results}
\begin{tabular}{@{}p{0.45\columnwidth}lcr@{}}
\toprule

\textbf{Model} & \textbf{Method} & \textbf{Acc~(\%)} & \textbf{Cost} \\
\midrule

Qwen2.5-14B-Instruct 
& DyLAN 
& 52.64 
& 13926.8 \\

\rowcolor{rowhl}
Qwen2.5-14B-Instruct 
& \modelname{} 
& \textbf{63.28} 
& \textbf{2966.4} \\

\midrule

DS-R1-Distill-Qwen-14B 
& DyLAN 
& 65.91 
& 49780.3 \\

\rowcolor{rowhl}

DS-R1-Distill-Qwen-14B 
& \modelname{} 
& \textbf{69.94} 
& \textbf{14241.7} \\
\bottomrule
\end{tabular}
\end{table}

\subsection{Cost of Training the Orchestration Policy}
\label{app:rl-training-cost}

We use GRPO-style group sampling with group size $G=8$. The orchestration policy is optimized using AdamW with learning rate $1\times 10^{-5}$. The cost penalty coefficient is set to $\beta=0.02$. The maximum sequence length is 4096. We run \modelname{} on $8\times$ A800 GPUs. The policy optimization stage takes approximately 12--24 hours per dataset, depending on dataset size and convergence behavior. Including structural-prior learning, the total training cost is approximately 15--30 hours per dataset. This cost is incurred once per dataset. After training, \modelname{} amortizes this cost by pruning redundant agent invocations and producing more efficient coordination trajectories. 

\begin{table}[H]
\centering
\footnotesize
\setlength{\tabcolsep}{3pt}
\caption{Training cost and inference-time efficiency.}
\label{tab:rl_training_cost}
\resizebox{\columnwidth}{!}{
\begin{tabular}{lccr}
\toprule
\textbf{Method} & \textbf{Extra Training} & \textbf{Train Time} & \textbf{Cost / Acc~(\%)} \\
\midrule
Vanilla LLM & No & 0 hrs & 251.3 / 80.54 \\
DyLAN & No & 0 hrs & 12784.6 / 84.37 \\
MaAS & No & 0 hrs & 3625.7 / 85.98 \\

\midrule
\rowcolor{rowhl}
\modelname{} & Yes & 12--24 hrs/dataset & \textbf{2848.9 / 86.89} \\
\bottomrule
\end{tabular}
}
\end{table}

\noindent As shown in Table~\ref{tab:rl_training_cost}, although \modelname{} introduces additional one-time training cost, it achieves the best accuracy while maintaining lower inference cost than multi-agent baselines. The policy learns reusable orchestration behavior during training and reduces unnecessary agent calls during inference.

\subsection{Case Study}
\label{app:agent-config-orchestration}

\paragraph{Task-Conditioned Agent Specialization.}
As shown in Table~\ref{tab:agent_relevance_scores}, \texttt{AlgorithmDesigner} and \texttt{CodeWriting} receive different relevance scores because their prompts encode distinct functional roles: \texttt{AlgorithmDesigner} focuses on designing the solution before implementation, whereas \texttt{CodeWriting} focuses on producing executable code. Thus, even within the same broad domain, agents can differ in their task-level relevance.

\begin{table*}[!t]
\centering
\small
\setlength{\tabcolsep}{5pt}
\renewcommand{\arraystretch}{1.15}
\caption{Examples of task-conditioned agent relevance scores.}
\label{tab:agent_relevance_scores}
\begin{tabular}{lcccccc}
\toprule
\textbf{Task} 
& \textbf{AlgorithmDesigner} 
& \textbf{CodeWriting} 
& \textbf{CodeReviewer} 
& \textbf{UnitTestWriter} 
& \textbf{EdgeCaseHunter} 
& \textbf{BugFixer} \\
\midrule
below\_zero & 0.52 & 0.86 & 0.48 & 0.71 & 0.41 & 0.39 \\
find\_zero  & 0.66 & 0.88 & 0.61 & 0.73 & 0.58 & 0.54 \\
\bottomrule
\end{tabular}
\end{table*}

Importantly, node-level relevance only determines which agents are retained after filtering; the final orchestration is not determined by relevance alone. \modelname{} further combines node-level relevance with edge-level transition weights in Eq.~\ref{eq:graphspec}. Therefore, even when two agents are both relevant to the same domain, their coordination patterns may differ depending on interaction plausibility and task-instance complexity.

\begin{table*}[t]
\centering
\footnotesize
\setlength{\tabcolsep}{3pt}
\caption{Case comparison between manual orchestration and \modelname{} on HumanEval.}
\label{tab:manual_vs_maca_cases}

\begin{tabular}{@{}p{0.12\textwidth}p{0.10\textwidth}p{0.58\textwidth}p{0.07\textwidth}p{0.07\textwidth}@{}}
\toprule
\textbf{Task} & \textbf{Method} & \textbf{Orchestration Path} & \textbf{Result} & \textbf{Cost} \\
\midrule

\texttt{below\_zero}
& Manual
& \texttt{AlgorithmDesigner} $\rightarrow$ \texttt{CodeWriting} $\rightarrow$ \texttt{UnitTestWriter} $\rightarrow$ \texttt{BugFixer}
& Pass
& 1.63k \\

\rowcolor{rowhl}
\texttt{below\_zero}
& \modelname{}
& \texttt{CodeWriting} $\rightarrow$ \texttt{UnitTestWriter} $\rightarrow$ \texttt{STOP}
& Pass
& 0.84k \\

\midrule

\texttt{find\_zero}
& Manual
& \texttt{AlgorithmDesigner} $\rightarrow$ \texttt{CodeWriting} $\rightarrow$ \texttt{UnitTestWriter} $\rightarrow$ \texttt{BugFixer}
& Fail
& 2.41k \\

\rowcolor{rowhl}

\texttt{find\_zero}
& \modelname{}
& \texttt{AlgorithmDesigner} $\rightarrow$ \texttt{CodeWriting} $\rightarrow$ \texttt{EdgeCaseHunter} $\rightarrow$ \texttt{CodeReviewer} $\rightarrow$ \texttt{BugFixer} $\rightarrow$ \texttt{UnitTestWriter} $\rightarrow$ \texttt{STOP}
& Pass
& 3.54k \\

\bottomrule
\end{tabular}
\end{table*}

\begin{table*}[!t]
\centering
\small
\setlength{\tabcolsep}{3.5pt}
\caption{Comparison between LLM-router variants and \modelname{} on GSM-Hard.}
\label{tab:rl_vs_llm_router}

\begin{tabular}{@{}llccc@{}}
\toprule
\textbf{Orchestrator} 
& \textbf{Constraint} 
& \textbf{Optimizable} 
& \textbf{Acc~(\%)} 
& \textbf{Invalid Action Rate} \\
\midrule
LLM Router 
& None 
& \xmark 
& 42.42 
& 6.4\% \\

LLM Router + GraphSpec Prompt 
& Natural-language hint 
& \xmark
& 45.45 
& 3.5\% \\

LLM Router + Mask 
& Rule-based constraint 
& \xmark
& 47.47 
& 1.1\% \\

\midrule
\rowcolor{rowhl}

\modelname{} 
& Mask + Prior regularization 
& \cmark 
& \textbf{50.30} 
& \textbf{0.0\%} \\
\bottomrule
\end{tabular}

\end{table*}

\paragraph{Adaptive Orchestration Versus Manual Orchestration.}
To further illustrate the difference between fixed manual orchestration and adaptive orchestration, we compare \modelname{} with a manually designed workflow on two HumanEval examples in Table~\ref{tab:manual_vs_maca_cases}. Manual orchestration follows a fixed path, whereas \modelname{} adjusts the path according to task difficulty. For a simpler task, \modelname{} uses a shorter path and reduces token cost. For a harder task, it invokes additional verification and repair agents, improving the final result.

\subsection{RL-based Orchestration versus LLM Routers}
\label{app:rl_vs_llm_router}

We further compare the learned orchestration policy in \modelname{} against an LLM-based router. 
An LLM router relies on pretrained instruction-following behavior and remains static with respect to task-specific outcomes. 
In contrast, the RL policy in \modelname{} is optimized using task-level rewards, allowing the routing strategy to adapt to the target benchmark and budget constraints. Table~\ref{tab:rl_vs_llm_router} compares \modelname{} with several LLM-router variants on GSM-Hard. 
The plain LLM router achieves 42.42\% accuracy and exhibits a 6.4\% invalid action rate. 
Adding GraphSpec as a natural-language hint improves accuracy, while enforcing GraphSpec as a hard mask further reduces invalid actions. 
However, both variants remain non-optimizable and are still weaker than \modelname{}. 
With mask-based constraints and prior regularization, \modelname{} improves accuracy by 7.88\% over the plain LLM router and reduces invalid actions to zero, confirming the necessity of RL-based orchestration.

\end{document}